\documentclass[11pt]{article}
\usepackage{a4wide}
\usepackage{epsfig,epsf,graphics}
\usepackage{times}
\usepackage{color}
\usepackage{amsfonts,amssymb,latexsym,amsmath} 
\usepackage{bm}
\usepackage{enumerate}
\usepackage{enumitem}
\usepackage{hhline}
\usepackage{longtable}
\usepackage{array} 
\usepackage{epstopdf}

\newcommand{\email}[1]{\footnote{{\em E-mail address:} \texttt{#1}}}

\begin{document}
\title{Probing the nature of $Y(4260)$ and $Z_c(3900)$ in the isospin violating process $Y(4260) \to J/\psi \eta \pi^0$}

\author{Xiao-Gang Wu$^{\, a, b}$\email{wuxiaogang@ihep.ac.cn} ,
Christoph Hanhart$^{\,b,}$\email{c.hanhart@fz-juelich.de} ,
Qian Wang$^{\, b,}$\email{q.wang@fz-juelich.de} ,
Qiang Zhao$^{\, a,}$\email{zhaoq@ihep.ac.cn} \\ 
  {\it\small$^a$Institute of High Energy Physics and Theoretical Physics Center
for Science Facilities, } \\
   {\it\small  Chinese Academy of Sciences, Beijing 100049, China}\\
   {\it\small$^b$Institut f\"{u}r Kernphysik, Institute for Advanced
Simulation, and J\"ulich Center for Hadron
Physics,}\\
   {\it \small D-52425 J\"{u}lich, Germany} \\
    }

\maketitle

\begin{abstract}
The isospin violation process $Y(4260) \to J/\psi \eta \pi^{0}$ is studied
assuming that $Y(4260)$ is a $D_{1} \bar{D}+c.c.$ hadronic molecule.
In association with the production of the $Z_c(3900)$, which is treated as a $D \bar{D}^{*}+c.c.$ hadronic molecule,
this  process can help us distinguish their  molecular natures from other scenarios, since
 the incomplete cancellation between the charged and neutral--meson loops, which are prominent
 in the molecular picture only,
  produces a peak in the $e^+e^-\to Y(4260)\to J/\psi\eta\pi^0$ cross section at the $D_{1} \bar{D}+c.c.$ threshold and
  a very prominent peak in  the $J/\psi \eta$ invariant mass spectrum in between the $D \bar{D}^{*}+c.c.$ thresholds; the latter being
much narrower than the corresponding one in the isospin conserving channel, i.e. $J/\psi \pi^+ \pi^{-}$.
     The partial width of $Y(4260)\to J/\psi\eta\pi^0$ is about $4 \times 10^{-4}$ of that of $Y(4260)\to J/\psi\pi^+\pi^-$. The cross section of $e^+e^-\to Y(4260)\to J/\psi\eta\pi^0$  at the $D_{1} \bar{D}+c.c.$ threshold is about $0.05 \ \mathrm{pb}$ which is much larger than that produced by the nearby resonances. These features are the direct consequences  of the assumed nature of these two states which might be accessible at the high-statistics experiments such as BESIII and LHCb.
 \end{abstract}

\newpage

\section{Introduction}
Early this year, the BESIII Collaboration reported a new charged charmonium-like structure $Z_{c}(3900)^{\pm}$ in the $J/\psi \pi^{\pm}$ invariant mass spectrum in the reaction of $Y(4260) \to J/\psi \pi^{+} \pi^{-}$~\cite{Ablikim:2013mio}. This result was soon confirmed by the Belle Collaboration~\cite{Liu:2013dau} and in an analysis based on data from the CLEO-c experiment~\cite{Xiao:2013iha}. In addition, Ref.~\cite{Xiao:2013iha} also reported the neutral $Z_{c}^{0}(3900)$ at a $3\sigma$ significance level  in $e^{+} e^{-} \to J/\psi \pi^{0} \pi^{0}$ at $\sqrt{s}=4170 \ \text{MeV}$. These observations immediately initiated a lot of theoretical studies of the $Z_c(3900)$ based on different scenarios such as hadronic molecule~\cite{Wang:2013cya,Guo:2013sya,Wilbring:2013cha}, tetraquark~\cite{Faccini:2013lda}, hadro-charmonium~\cite{Voloshin:2013dpa} and threshold effects~\cite{Wang:2013cya,Chen:2013coa}.

The pole of the $Z_c(3900)$ is located near the $D\bar{D}^{*}$ \footnote{We implicitly include the charge conjugation state here and below. The case is the same for $D_1\bar D$.} thresholds and the $Y(4260)$ is near the $S$-wave $D_{1}\bar{D}$ thresholds. It was proposed that the $Y(4260)$
and the $Z_c(3900)$ could have a sizeable $D_{1}\bar{D}$ and $D\bar D^*$ component,
respectively~\cite{Wang:2013cya}. In this scenario the $Y(4260)$ first couples
to $D_{1}\bar{D}$ in an $S$-wave followed by the $D_{1}$ decay to $D^{*}\pi$ in a $D$-wave.
Then the strong interactions between the low momentum $D$ and $\bar{D}^*$ will produce the $Z_{c}(3900)$
near threshold. In Ref.~\cite{Wang:2013cya} based on the above picture,  the invariant mass spectra of $Y(4260) \to J/\psi \pi^{+} \pi^{-}$
were analyzed, where contributions from both the triangle diagrams with an explicit $Z_c(3900)$ pole and box diagrams without the pole were
 considered. It turned out that both  contributions are needed to reproduce the $J/\psi \pi^{+}$ and $\pi^{+}\pi^{-}$ mass spectra with
a clear dominance of the box diagrams. Besides this explanation, various other suggestions
exist for the nature of $Y(4260)$,
such as the conventional $4 S$ charmonium state~\cite{LlanesEstrada:2005hz},
hadro-charmonium~\cite{Voloshin:2007dx,Dubynskiy:2008mq,Li:2013ssa},
hybrid~\cite{Zhu:2005hp,Kou:2005gt,Close:2005iz},
$\chi_{c0}\omega$ molecule state~\cite{Dai:2012pb}
and tetraquark state~\cite{Maiani:2005pe}.  In order to further constrain the reaction dynamics and to gain deeper insights into the nature of
both $Y(4260)$ and $Z_{c}(3900)$, we propose in this work to investigate the isospin violating process $Y(4260) \to J/\psi \eta \pi^{0}$.
We will show that the incomplete cancellation between the charged and neutral charmed-meson loops can produce
a peak in the $e^+e^-\to
Y(4260)\to J/\psi\eta\pi^0$ cross section at the $D_{1} \bar{D}$ threshold and a very narrow peak in  the $J/\psi \eta$ invariant mass spectrum
 at the $D \bar{D}^{*}$ threshold, which is much more significant and narrower than that in the isospin conserving channel, i.e. $J/\psi \pi^0 \pi^{0}$ since the width is given by the distance
 of the charged to neutral $D\bar D^*$ thresholds.
 We argue that these are distinct features of the molecular scenario.

The study of the isospin violation process has several benefits.
Firstly, the isospin violation process is usually clean compared to the isospin conserved process. The background is reduced significantly
which will make it much easier to identify an isospin violating structure. For instance, in Ref.~\cite{Wang:2011yh} it has been shown that the open charm effects may be easily identified in the isospin violation process $e^{+}e^{-} \to (c\bar{c})_{1^{--}} \to J/\psi \pi^{0}$ in contrast to those in the isospin conserved processes $e^{+}e^{-} \to J/\psi \eta , \phi \eta_{c}$.  So we expect that the background in $Y(4260) \to J/\psi \eta \pi^{0}$ is simpler than that in $Y(4260) \to J/\psi \pi^{+} \pi^{-}$.
Secondly, the isospin violation process may be enhanced by loop effects, especially when molecular structures
are involved. As an example, BESIII recently reported the anomalously large isospin violations in
 $J/\psi \to \gamma \eta(1405/1475) \to \gamma 3\pi$~\cite{BESIII:2012aa}. This phenomenon was explained later by a
 triangle singularity mechanism~\cite{Wu:2011yx,Wu:2012pg,Aceti:2012dj}. It shows that the hadron loops may cause
  larger isospin violation effects than the direct mixing in threshold production processes. Another example is the very narrow structure
observed in $J/\psi\to \phi \pi^0\eta$~\cite{a0f0mixing_exp} predicted to occur in Refs.~\cite{a0f0mixing_th1,a0f0mixing_th2}. Here the width
of the structure is determined by the distance between the charged and neutral kaon pair thresholds which enters via the kaon loops.
It is also worth mentioning the study of  the hadronic width
of the $D_{s0}^*(2317)$ in the context. In Refs.~\cite{Faessler:2007gv,Lutz:2007sk,Guo:2008gp}, it was shown that due to the
same interplay of loop contributions the hadronic width of the $D_{s0}^*(2317)$ from its isospin
violating decay to $\pi^{0}D_{s}$ gets enhanced significantly, if the $D_{s0}^*(2317)$ is a $DK$
molecule.
This is exactly  analogous to
the mechanisms at work in this paper. On the other hand, the structure would have been a lot broader, if any other mechanisms  but the kaon loops (for instance, the $\pi^0-\eta$ mixing)
had been dominant.  Thus, in this work we study the effect of both the heavy meson loops as well as $\pi^0-\eta$ mixing.
Thirdly, in the charmonium energy region,  we have several high-statistics and high-luminosity machines feeding experiments
 such as BESIII, Belle, BaBar and LHCb. It is very promising that the isospin violation process $Y(4260) \to J/\psi \eta \pi^{0}$  can be accessible at one of these facilities.

This work is organized as the following: We illustrate our framework in Sec.~\ref{sec:Formalism}. The results and discussions are presented in Sec.~\ref{sec:result} and a summary is given in the last section.

\section{Framework}\label{sec:Formalism}

\begin{figure}[t!]
\begin{center}
\includegraphics[width=0.8\linewidth]{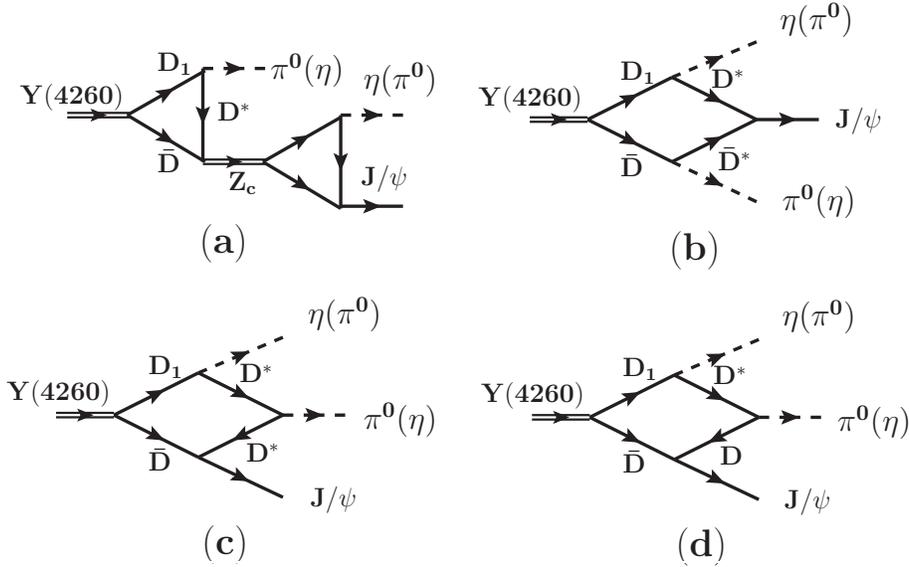}
\caption{Feynman diagrams for $Y(4260) \to J/\psi \eta \pi^{0}$.}
\label{fig:1}
\end{center}
\end{figure}

\subsection{Feynman Diagrams}
\begin{figure}[t!]
\begin{center}
\includegraphics[width=0.3\linewidth]{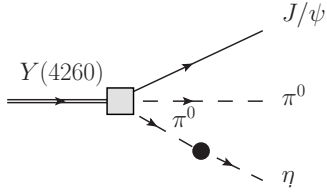}
\caption{The schematic diagram for the $\pi^{0}$-$\eta$ mixing. The square denotes that all the allowed diagrams in Fig.~\ref{fig:1} contributing to $Y(4260)\to J/\psi \pi^0\pi^0$ are included and the round dot represents the mixing between $\pi^0$ and $\eta$. }
\label{fig:Mixing}
\end{center}
\end{figure}

Our calculation is based on the assumption that the $Y(4260)$ is dominated by $D_1\bar D$ and $Z_c(3900)$ is dominated by $D\bar D^*$. The relevant Feynman diagrams for  $Y(4260) \to J/\psi \eta \pi^{0}$ are listed in Fig.~\ref{fig:1}, where Fig.~\ref{fig:1}(a) represents the triangle diagrams through an intermediate $Z_{c}(3900)$ and Fig.~\ref{fig:1}(b-d) are box diagrams which are similar to those in Ref.~\cite{Wang:2013cya} for the isospin conserving process $Y(4260) \to J/\psi \pi^0 \pi^{0}$. Apart from these contributions, we also consider the contribution from $\pi^{0}-\eta$ mixing as shown in Fig.~\ref{fig:Mixing}. The gray square means that all the possible diagrams of $Y(4260) \to J/\psi \pi^0 \pi^{0}$ as that in Ref. ~\cite{Wang:2013cya} are included and the black circle is the mixing between $\pi^0$ and $\eta$. The mixing intensity can be determined by
\begin{eqnarray}\label{eq:8}
\epsilon_{0}=\frac{1}{\sqrt{3}} \frac{ M_{K^{0}}^{2} - M_{K^{+}}^{2} + M_{\pi^{+}}^{2} - M_{\pi^{0}}^{2} }{ M_{\eta}^{2} - M_{\pi^{0}}^{2} } = 0.01
\end{eqnarray}
using Dashen's theorem~\cite{Dashen:1969eg}.

\subsection{Effective Lagrangians}
The diagrams in our framework are described by the NREFT introduced in Refs.~\cite{Fleming:2008yn,Guo:2010ak} in which  the heavy fields
are treated nonrelativistically while the light mesons, $\pi$ and $\eta$, are treated relativistically.

Although the related Lagrangians and couplings in this paper can be found in Refs.~\cite{Wang:2013cya,Guo:2013zbw}, we list some of them for completeness. By assuming $Y(4260)$ to be an $S$-wave $D_1 \bar{D}$ molecular state with $I^G(J^{PC})=0^-(1^{--})$,  the Lagrangian for its coupling to the constituents reads~\cite{Wang:2013cya,Guo:2013zbw}
\begin{equation}\label{eq:1}
\mathcal{L}_{Y}=\frac{y}{\sqrt{2}} Y^{i\dag}(D_{1a}^{i} \bar{D}_{a} - D_{a} \bar{D}_{1a}^{i}) + h.c.
\ ,
\end{equation}
where $Y^{i\dag}$ is the creation operator for $Y(4260)$ and the other operators denote the annihilation operators for the corresponding particles.
 The renormalized effective coupling $y$ is related to the probability of finding $D_{1} \bar{D}$ component in the physical wave function of $Y(4260)$ which can be estimated from Weinberg's compositeness theorem~\cite{Weinberg:1965zz,Baru:2003qq}. Based on these considerations
 $|y|=3.28\pm{1.4}  \  {\textrm{GeV}}^{-1/2},$
  is extracted in Refs.~\cite{Wang:2013cya,Guo:2013zbw}. However,
  all results shown here are insensitive to the value of $y$, since the overall normalization of the predicted rates is fixed by the
  measured $e^+e^-\to J/\psi\pi\pi$ cross section.

The newly discovered $Z_c(3900)$~\cite{Ablikim:2013mio,Liu:2013dau,Xiao:2013iha} is a charged charmonium-like state with $I^{G}(J^{PC})=1^{+}(1^{+-})$ for its charge-neutral state which is the charm sector's analogue of $Z_{b}$. If $Z_c(3900)$ is an $S$-wave $D \bar{D}^{*}$ molecular state, the interaction Lagrangian reads~\cite{Cleven:2013sq}
\begin{eqnarray}\label{eq:3}
\mathcal{L}_{Z}= z ( \bar{V}^{\dag i} Z^{i} P^{\dag} - \bar{P}^{\dag} Z^{i} V^{\dag i} )+h.c.
\end{eqnarray}
The $Z_c(3900)$  isotriplet can be written as a $2\times2$ matrix
\begin{equation}\label{eq:4}
Z_{ba}=\left( \begin{array}{cc}
            \frac{1}{\sqrt{2}} Z^{0} & Z^{+} \\
            Z^{-}       &  -\frac{1}{\sqrt{2}} Z^{0}
            \end{array}  \right)_{ba}
\ ,
\end{equation}
and the charmed mesons are given by $P(V)=(D^{(*)0},D^{(*)+})$.
Current data does not allow one to decide, if the $S$-matrix singularity related to the $Z_c(3900)$ is located above
or below the $D\bar D^*$ threshold and thus we cannot calculate the parameter $z$ analogously to
what was done in case of the $Y(4260)$.
Phenomenologically, however, we can get this coupling constant from an analysis of the data
for $Y(4260)\to J/\psi\pi^+\pi^-$. The analysis of Ref.~\cite{Wang:2013cya} revealed
\begin{equation}\label{eq:5}
|z|=(0.77 \pm 0.23) \ \textrm{GeV}^{-1/2}
\ .
\end{equation}

To incorporate the $\eta$ meson, we adopt the pseudoscalar octet
\begin{equation}\label{eq:6}
\phi=
\left(\begin{array}{ccc}
\frac{1}{\sqrt{2}}\pi^{0} + \frac{1}{\sqrt{6} }\eta  & \pi^{-} & K^{+} \\
\pi^{-} & -\frac{1}{\sqrt{2}}\pi^{0} + \frac{1}{\sqrt{6} }\eta  & K^{0} \\
K^{-} & \bar{K}^{0} & -\frac{2}{\sqrt{6} }\eta
\end{array}\right)
\ ,
\end{equation}
where we have identified the $\eta$ meson as the SU(3) octet element $\eta_{8}$. In the heavy quark limit, the heavy and light degrees of freedom are conserved separately. So the heavy mesons can be classified by their light degrees of freedom, i.e., $s_{l}=s_{q}+l$ with $s_{q}$ the spin of the light quark and $l$ the orbital angular momentum. The narrow $P$-wave meson $D_{1}$ is considered as a $s_{l}=3/2$ state and decays to $D^{*} \pi$ in a $D$ wave. The interaction Lagrangian reads~\cite{Colangelo:2005gb}
\begin{small}
\begin{equation}\nonumber
\mathcal{L}_{D_{1}} = i \frac{h^{\prime}}{f_{\pi}} \left[
    3 D_{1a}^{i}(\partial^{i} \partial^{j} \phi_{ab}) D_{b}^{*\dag j} - D_{1a}^{i}(\partial^{j} \partial^{j} \phi_{ab}) D_{b}^{*\dag i} + 3\bar{D}_{a}^{*\dag i} (\partial^{i} \partial^{j} \phi_{ab}) \bar{D}_{1b}^{j} - \bar{D}_{a}^{*\dag i} (\partial^{j} \partial^{j} \phi_{ab}) \bar{D}_{1b}^{i}
\right] + h.c.
\ .
\end{equation}
\end{small}
From the width of $D_{1}$, the coupling $h^{\prime}$ is determined to be $|h^{\prime}|=(0.62 \pm 0.08) \ \textrm{GeV}^{-1}$~\cite{Cleven:2013mka}.

\subsection{$e^+e^-\to Y(4260)\to J/\psi\eta\pi^0$ and the propagator of $Y(4260)$}
The cross section of $e^+e^-$ annihilation to any final states via a vector meson can be expressed by the vector meson dominance via the effective photon-vector-meson coupling $g_{\gamma^* V}$ (see e.g.~\cite{VMD}). For the full process $e^+e^-\to Y(4260)\to J/\psi\eta\pi^0$, this yields
\begin{eqnarray}
\sigma(s)=(4\pi\alpha)^2\left (g_{\gamma^*Y}\frac{M_Y^2}{s}\right)^2(M_Y\Gamma_{Y\to J/\psi\eta\pi^0})|G_Y(s)|^2
\label{eq:crosssection}
\end{eqnarray}
where $g_{\gamma^*Y}$ is the dimensionless coupling constant between the virtual photon and vector state $Y(4260)$, and $G_Y(s)$ is the propagator of $Y(4260)$, i.e. ~\cite{Cleven:2013sq}
\begin{eqnarray}
G_Y^{-1}=s-M_Y^2+\hat\Pi\left(s\right)+iM_Y\Gamma_Y
\end{eqnarray}
with
\begin{eqnarray}
\hat\Pi\left(s\right)=\Pi\left(s\right)-\mathrm{Re}\left[\Pi\left(M_Y^2\right)+\left(s-M_Y^2\right)\partial_{s}\Pi(s)|_{s=M_Y^2}\right] \ .
\label{eq:DoubleSubtraction}
\end{eqnarray}
In the above equation, the self energy $\hat\Pi\left(s\right)$ is doubly-subtracted at mass position $M_Y=(4220\pm 5)~\mathrm{MeV}$ which is fitted by the data for $e^+e^-\to Y(4260)\to J/\psi\pi^+\pi^-$ and $h_c\pi^+\pi^-$~\cite{Cleven:2013mka}.  Here $\Pi(s)$ is the $D_1\bar D$ bubble diagram contributing to the $Y(4260)$ self energy.
 $\Gamma_Y=(40\pm 9)~\mathrm{MeV}$ is the constant partial decay width of the $Y(4260)$ without going through
  the $D_1\bar D$ component, namely, the non-$D_1\bar{D}$ decay width~\cite{Cleven:2013mka}.

\section{Results and discussion}\label{sec:result}
\subsection{$J/\psi\eta$, $J/\psi\pi^0$ and $\eta\pi^0$ invariant mass distributions}\label{subsec:invariantmass}

\begin{figure}[t]
\begin{center}
\includegraphics[width=1.\linewidth]{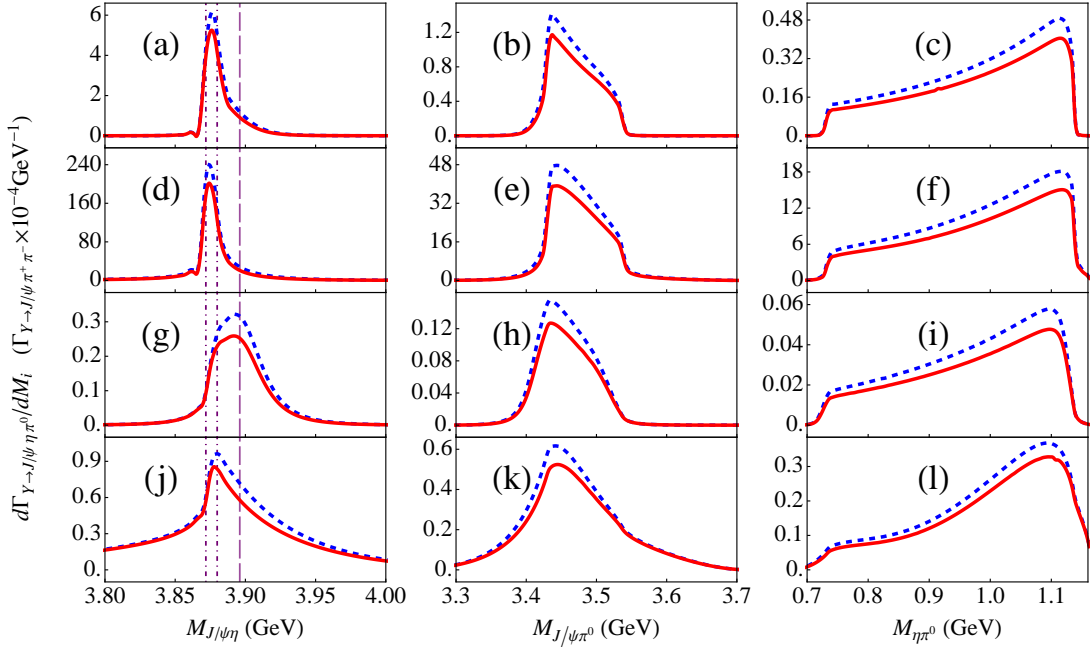} 
\caption{The invariant mass spectra for the final state $J/\psi\eta$ (left column), $J/\psi\pi^0$ (middle column) and $\eta\pi^0$ (right column) in $Y(4260) \to J/\psi \eta \pi^{0}$ evaluated for a total energy of 4260 MeV. The figures in the first, second, third and last row are the contributions from the triangle diagrams  (Fig.~\ref{fig:1}(a)), box diagrams (Fig.~\ref{fig:1}(b))), $\pi^{0}$-$\eta$ mixing  through triangle diagrams and $\pi^{0}$-$\eta$ mixing through box diagrams, respectively. The solid (red) and dashed (blue) lines stand for the contributions with and without considering the width of $D_1$.  The three vertical lines in the left column denote the $D^{0} \bar{D}^{*0}$ and $D^{+} \bar{D}^{*-}$ threshold, and the $Z_c(3900)$ mass, respectively. All the differential partial widths have been normalized to the partial width of the isospin conserving process, i.e. $\Gamma_{Y(4260)\to J/\psi\pi^+\pi^-}$. }
\label{fig:etapi}
\end{center}
\end{figure}

In Fig.~\ref{fig:etapi} the numerical results for $Y(4260) \to J/\psi \eta \pi^{0}$  are presented, where the invariant mass spectra for the final state $J/\psi\eta$, $J/\psi\pi^0$ and $\eta\pi^0$ are plotted for each mechanism individually. Since there are still large uncertainties with the coupling between $Y(4260)$ and $D_1\bar D$, it is more convenient to define the ratio of the partial width of  $Y(4260) \to J/\psi \eta \pi^{0}$ to $\Gamma_{Y(4260)\to J/\psi\pi^+\pi^-}$, in which case the coupling uncertainties cancel. As shown by the left column of Fig.~\ref{fig:etapi}, i.e. (a), (d), (g), and (j),  a peak around $3.9 \ \textrm{GeV}$ appears in the invariant mass spectrum of  $J/\psi \eta$ in all the cases. However, the line shapes
as well as the predicted rates are quite different from each other: The peaks from the triangle (Fig.~\ref{fig:etapi}(a)) and box diagrams (Fig.~\ref{fig:etapi}(d)) are located at the $D\bar{D}^{*}$ thresholds with a narrow width of about $8 \ \textrm{MeV}$ reflecting the mass difference between the charged and neutral $D\bar{D}^{*}$ thresholds. In addition, in these cases the peaks are asymmetric and the asymmetry is most pronounced
in Fig.~\ref{fig:etapi}(a) where the  $Z_c(3900)$ pole contributes.
The reason for the appearance of this narrow structure is that the charged and neutral meson loops interfere destructively in isospin violating
transitions, since these two amplitudes have to cancel in the absence of the $D(D^*)$ meson mass differences.
In effect the leading isospin violating contribution comes from the difference of the two-meson cut contributions of
the individual amplitudes which is proportional to the phase space and its analytic continuation to below
threshold  in the
neutral and charged channels, respectively. This contribution is
  non-analytic in the quark masses and can lead to significantly enhanced violating effects, namely, the sum of  amplitudes driven by the meson
  mass differences will outnumber those from $\pi^0-\eta$ mixing by one order of magnitude.
  The same mechanism is also responsible for the isospin violation decay $\eta(1405) \to 3 \pi$~\cite{Wu:2011yx,Wu:2012pg}
  and $a_0-f_0$ mixing in $J/\psi\to \phi\pi^0\eta$~\cite{a0f0mixing_th1,a0f0mixing_th2}.
  For those two cases, since their decay mechanism is through the kaon loops, a similar peak near the
  $K\bar K^{(*)}$ threshold has been found in $\pi\pi$ and $\eta\pi^0$ invariant mass spectra.  It shows
  that the dominant isospin violating process is sensitive to the
  significance of the
  intermediate two-meson states in the wave functions of the relevant hadron. 
  In this sense it provides a direct measure of the molecular component of the states.

On the other hand, the spectra from any mechanism that is not driven by the mass differences of the open charm mesons in the loops
are expected to be significantly broader and more similar to the isospin conserving counter parts.
For illustration in Fig.~\ref{fig:etapi}(g) and (j) we show the spectra that emerge, if the isospin violation comes from  $\pi^{0}$-$\eta$ mixing.
The reason is that here the transition matrix element is the same as its isospin conserving counter parts, since
 the isospin violation occurs only on one of the external pion legs.
Especially, in these two cases the loops interfere constructively.

For the $J/\psi \pi^{0}$ spectrum, there is a broad bump near $3.4-3.5 \ \textrm{GeV}$ in all the four mechanisms as shown in the middle column of Fig.~\ref{fig:etapi}, i.e. (b), (e), (h), and (k),  which is the kinematical reflection of the narrow peak in the $J/\psi \eta$  spectrum. The line shape of the $\eta \pi^{0}$ spectrum is very similar to the $\pi^{+} \pi^{-}$ spectrum investigated in $Y(4260)\to J/\psi\pi^+\pi^-$ before considering the finial state interactions. The structures near $0.73\ \textrm{GeV}$ and $1.13\ \textrm{GeV}$ are also the kinematical reflections of the peak in the $J/\psi \eta$ spectrum~\cite{Wang:2013cya}.

For the $J/\psi \pi^{+} \pi^{-}$ channel the BESIII data~\cite{Ablikim:2013mio} provide a constraint on the ratio
between the triangle and box diagrams ~\cite{Wang:2013cya},\footnote{Here we do not  consider the $\pi\pi$ final state interaction which in the scenario discussed here only
gives a small correction since the pions are predominantly in a $D$ wave.}
\begin{equation}\label{eq:9}
\frac{ \Gamma(Y(4260)  \to J/\psi \pi^{+} \pi^{-})_{\text{triangle}} } { \Gamma(Y(4260)  \to J/\psi \pi^{+} \pi^{-})_{\text{box}}  } \approx 12\%
\ .
\end{equation}
In case of the isospin violating the dominance of the box diagram is even larger: the analogous ratio to
Eq.~(\ref{eq:9}) here gives 2\%.
In order to better understand
how the isospin violation works quantitatively for each mechanism individually, we define the ratio
\begin{equation}\label{eq:9a}
\xi_{m}\equiv\frac{\Gamma(Y(4260)  \to J/\psi \eta \pi^{0})_{m}}{\Gamma(Y(4260)  \to J/\psi \pi^{+} \pi^{-})_{m}}
\ ,
\end{equation}
where the subscript represents the specific mechanism. 
It should be stressed that the values of the ratio $\xi_m$ cannot be compared
directly to observables, since the quantity in the denominator provides in general only a small fraction of the cross section.

One finds that the box contribution with the isospin violation driven by the meson mass differences
in the loop provides the largest effect, i.e. $\xi_{\rm box}=4\times 10^{-4}$. In contrast,  the triangle diagrams give in connection
with the same mechanism
 $\xi_{\rm triangle}=1 \times 10^{-4}$. The reason for this difference is that
the contribution from $Y(4260) \to Z_{c}^{0} \pi^{0} \to (J/\psi \eta) \pi^{0}$ is much larger than that from $Y(4260) \to Z_{c}^{0} \eta \to (J/\psi \pi^{0}) \eta$,
since the first triangle loop of the former process satisfies the triangle singularity condition~\cite{Wu:2011yx,Wu:2012pg,Wang:2013hga}, while the latter process does not.

On the other
hand, the two ratios for the diagrams where the isospin violation is modeled by the $\pi^0$-$\eta$ mixing are of similar size,
about $1\times 10^{-5}$, but a factor
of 40 smaller than the leading ratio $\xi_{\rm box}$.
The size of these ratios can be understood quantitatively, since the isospin violation can be estimated as via the difference in phase
spaces
\begin{equation}\label{eq:10}
\xi_{\text{mixing}}=
|\epsilon_{0}|^{2} \frac{P.S.(Y(4260) \to J/\psi \eta \pi^{0})}{P.S.(Y(4260) \to J/\psi \pi^{+} \pi^{-})} = 3.52 \times 10^{-5}
\ .
\end{equation}
This implies that all mechanisms that are not enhanced by the non-analytic isospin violating terms from the loops should be similar in size.

 As one can see from Fig.~\ref{fig:etapi} the width effects of the $D_{1}$ are about $10\%$ for each individual spectrum.
The main uncertainty in our calculation comes from the mass difference between the charged and neutral $D_{1}$,
for which we have adopted $M_{D_{1}^{+}}-M_{D_{1}^{0}}=M_{D^{*+}}-M_{D^{*0}}$. If we use an equal mass for the charged and neutral $D_{1}$, the results will change by about $30\%$ in those contributions where the isospin violation was driven
by the meson mass differences, while the changes to the mixing diagrams are negligible.
This is again because the  charged and neutral loops interfere destructively in the former group,
 while the interference is constructive in the latter.

\subsection{The line shape of $Y(4260)$ in $e^+e^-\to Y(4260)\to J/\psi\eta\pi^0$ process}

\begin{figure}[t!]
\begin{center}
\includegraphics[width=0.6\linewidth]{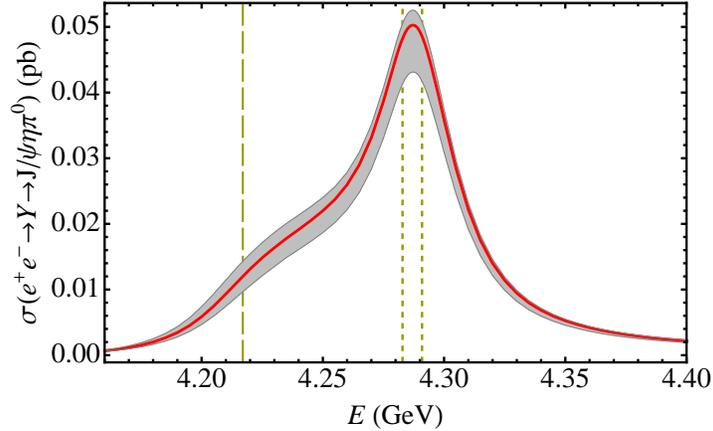} 
\caption{The predicted cross section for $e^+e^-\to Y(4260)\to J\psi\eta\pi^0$.
 The two vertical short-dashed lines denote the charged
and neutral $D_1\bar D$ thresholds, respectively, while the long-dashed line
denotes the location of the $Y(4260)$ pole as it emerged from the fit
to the isospin conserving data. The grey band shows the variation of our prediction,
when the parameters are allowed to vary within the statistical uncertainty allowed
by the fit to the isospin conserving data.}
\label{fig:JpsiEtaPi}
\end{center}
\end{figure}

Using Eq.~(\ref{eq:crosssection}) and the parameters fitted in $J/\psi\pi^+\pi^-$ and $h_c\pi^+\pi^-$ channels \cite{Cleven:2013mka}, we can
predict the cross section for $e^+e^-\to Y(4260)\to J/\psi\eta\pi^0$
as a function of the center of mass energy as shown in Fig.~\ref{fig:JpsiEtaPi}. Again, because of the destructive interference
 between the charged and neutral charmed-meson loops in the dominant contributions,
 the isospin-violating cross section is maximal, of order 0.05 pb, close to the charged and neutral $D_1\bar D$ thresholds and $not$
 at the location of the $Y(4260)$ pole. This is a highly non-trivial prediction of the scenario explored in this paper.
 Contrary to the spectra discussed in the previous subsection, the line shape of  $e^+e^-\to Y(4260)\to J/\psi\eta\pi^0$
 turns out to be very sensitive to the $D_1$ width, because the effect of the $D_1\bar D$ cut gets weakened
 by the fact that the width pushes the corresponding branch point into the complex plane. This results in
 the width of the resulting structure not being given by the splitting between the charged and neutral $D_1\bar D$ thresholds,
 but by the width of the $D_1$. To be more specific, when we switch off the $D_1$ width in our calculation the peak at the $D_1\bar D$
 thresholds gets very narrow and at the same time the predicted cross section between the thresholds rises
 by one order of magnitude.

It is interesting to compare the isospin violating mechanisms based on different scenarios.
In the tetraquark picture~\cite{Faccini:2013lda,Maiani:2005pe}, the isospin violating effect is included by inserting the explicit effective Lagrangian to the isospin symmetric amplitudes. In the hadro-charmonium picture~\cite{Voloshin:2007dx,Dubynskiy:2008mq,Li:2013ssa}, this breaking can only happen through the $\pi^0$-$\eta$ mixing. Both these two scenarios give relatively small cross sections that are proportional to $(m_d-m_u)/m_s$
 and especially are not related to the $D_1\bar D$ threshold.
Meanwhile, if $Y(4260)$ is a $\chi_{c0}\omega$ molecule~\cite{Dai:2012pb}, its main decay mode should be $\chi_{c0}+\omega$ or $\chi_{c0}+3\pi$ and the corresponding isospin violation decay channels would be $\chi_{c0}+\rho^0$ or $\chi_{c0}+2\pi$ through $\omega$-$\rho$
mixing.  This mixing strength is much smaller than that between $\pi^0$ and $\eta$. For instance, the recent analysis
of Ref.~\cite{myff} gave for the mixing strength a value of $-0.002$. This is a factor
of 5 smaller than the value given in Eq.~(\ref{eq:8}). The total strength of the cross section from this scenario would thus
be smaller
by a factor of more than three orders of magnitude compared to that in the $D_1\bar D$ molecule picture.
In addition, the $\chi_{c0}\omega$ molecule nature will be sensitive to the $\chi_{c0}\omega$ threshold instead of the $D_1\bar D$ threshold.   Therefore, a  measurement of a $D_1\bar D$ threshold enhancement in the isospin violation process
 would be an unambiguous proof for a prominent $D_1\bar D$ molecular nature of
the $Y(4260)$.

\section{Summary}\label{sec:summary}

In this work, we assume that $Y(4260)$ and $Z_{c}(3900)$ are hadronic molecules composed of $D_{1} \bar{D}$ and $D \bar{D}^{*}$, respectively, as in Refs.~\cite{Wang:2013cya,Guo:2013sya,Guo:2013zbw}. We investigate the isospin violation process $Y(4260) \to J/\psi \eta \pi^{0}$ by considering triangle diagrams, box diagrams and $\pi^{0}$-$\eta$ mixings. We find that the position and width of the $D\bar{D}^*+c.c.$ threshold peak in the $J/\psi \eta$ invariant mass spectrum depends on the production mechanism.
Within the scenario outlined in the paper we predict a very narrow peak (width below 10 MeV) located between the thresholds for the neutral and charged $D\bar D^*$ channels.
On the other hand, if the $Z_c$ is predominantly non-molecular, we
predict the appearance of a peak with a width about $46 \ \textrm{MeV}$ in the $J/\psi \eta$ spectra. The partial width of $J/\psi \eta \pi^{0}$ channel is about $4\times 10^{-4}$ with respect to that of the $J/\psi \pi^{+} \pi^{-}$ channel.  In addition, we predict that,
if the $Y(4260)$ is predominantly a $D_1\bar D$ molecule,
the line shape of $e^+e^-\to J/\psi \eta \pi^{0}$ is very different to that in the isospin conserving transition $e^+e^-\to J/\psi \pi \pi$. 
Especially, it should peak at the $D_1\bar D$ thresholds instead of the pole position of the $Y(4260)$.

It should be stressed that what was assumed in this paper is that the $Y(4260)$ and $Z_c(3900)$ are pure $D_1\bar D$ and $D\bar D^*$ molecules,
respectively. Any admixture of
other components in the wave functions would make the peak shown in Fig.~\ref{fig:JpsiEtaPi} smaller. Therefore,
future experimental investigations of the peculiar structures predicted in the $J/\psi\eta$ invariant mass  near the $D\bar D^*$  threshold and the
total cross section cross section near the $D_1\bar D$ threshold will be important steps towards an understanding of the nature of
both the $Y(4260)$ and the $Z_{c}(3900)$.

\section*{Acknowledgments}
The authors are in debt to Feng-Kun Guo for having benefitted from his ``AmpCalc.m" package  in our computation. This work is supported, in part,
by the National Natural Science Foundation of China (Grant Nos.
11035006 and 11121092), the Chinese Academy of Sciences
(KJCX3-SYW-N2), the Ministry of Science and Technology of China
(2009CB825200), and DFG and NSFC funds to the Sino-German CRC 110.

\end{document}